\documentclass[]{interact}

\usepackage{graphicx}
\usepackage{url}
\usepackage{color}

\usepackage{epstopdf}
\usepackage[caption=false]{subfig}

\usepackage[numbers,sort&compress]{natbib}
\bibpunct[, ]{[}{]}{,}{n}{,}{,}
\makeatletter
\def\NAT@def@citea{\def\@citea{\NAT@separator}}
\makeatother

\theoremstyle{plain}

\theoremstyle{definition}

\theoremstyle{remark}

\begin{document}


\title{{Clamp cell with \emph{in situ}} pressure monitoring for low-temperature neutron scattering measurements}

\author{
\name{A.~Podlesnyak\textsuperscript{a}\thanks{CONTACT A.~Podlesnyak. Email: podlesnyakaa@ornl.gov}, M.~Loguillo\textsuperscript{a}, G.~M. Rucker\textsuperscript{a}, B.~Haberl\textsuperscript{a}, R.~Boehler\textsuperscript{a}, G.~Ehlers\textsuperscript{b},  L. L.~Daemen\textsuperscript{a}, D. Armitage\textsuperscript{a}, M. D. Frontzek\textsuperscript{a} and M.~Lumsden\textsuperscript{a}}
\affil{\textsuperscript{a}Neutron Scattering Division, Oak Ridge National Laboratory, Oak Ridge, Tennessee 37831, USA}
\affil{\textsuperscript{b}Neutron Technologies Division, Oak Ridge National Laboratory, Oak Ridge, Tennessee 37831, USA}
}

\maketitle

\begin{abstract}
A clamp pressure cell for neutron scattering experiments at low temperatures and in external magnetic fields under pressure up to 2~GPa has been fabricated and tested. The cell provides optical access to the sample space that allows instantaneous pressure determination during sample loading, cooling and measuring using ruby and/or samarium doped strontium tetraborate fluorescence monitoring.
A new calibration curve of the pressure-induced shift of the $^7D_0 - ^5F_0$ (0-0) line in the fluorescent spectrum of SrB$_4$O$_7$:Sm$^{2+}$ for moderate pressures, $P\leqslant{2}$~GPa, is given.
\end{abstract}

\begin{keywords}
neutron scattering, pressure measurements; clamp cell; fluorescence.
\end{keywords}

\section{Introduction}

High-pressure investigations by means of neutron scattering give valuable information about crystal and magnetic structure transformations, as well as lattice and magnetic dynamics of the materials under study.
Pressure, as an external parameter, can tune quantum fluctuations and causes quantum phase transitions, in contrast to a classical phase transition induced by thermal fluctuations.
Pressures up to $\sim{90}$~GPa have been reached in neutron diffraction experiments using state-of-the-art diamond anvil cells (DAC)~\cite{Boehler,Guthrie10552,Boehler2017}.
However, alternative techniques for applying pressure in a more moderate range, $P\leqslant{2}$~GPa, are still widely used in many areas of solid and soft matter physics, chemistry and biology, simply because pressure in this range still has a large impact on many physical properties of matter.
Examples include, but are not limited to, unconventional superconductivity~\cite{Tafti,Chen}, exotic quantum states and quantum phase transitions~\cite{Luo13520,Fujiwara,Li}, colossal magnetoresistance~\cite{Cai}, insulator-metal transitions~\cite{Cheng1670}, spin crossover~\cite{Kengo}, and others.
The study of many of these physical phenomena demands cryogenic temperatures and high magnetic fields.
Therefore, pressure cells with pressures up to $\sim{2}$~GPa and suitable for neutron scattering techniques in combination with low temperatures and external magnetic fields open up a wide experimental area and are an essential tool for condensed matter studies.

Pressures up to 2~GPa can be achieved in neutron scattering using piston-cylinder (or clamp) type pressure cells.
High pressure neutron-diffraction and especially inelastic neutron scattering (INS) measurements traditionally require large sample volumes.
In a clamp cell the sample volume can be as large as 1~cm$^3$, that is much larger than in a DAC (typically less than 1~mm$^3$)~\cite{Klotz_book}.
On the other hand, the clamp cell can obviously reach higher pressures than a gas pressure cell (typically  less than 0.8~GPa).
One of the main drawbacks of the clamp cell is that the sample is loaded, pressurized and locked at room temperature outside the neutron instrument.
An accurate pressure monitoring at all stages of a measurement, and not only at sample loading time, is vital for a successful experiment.
A traditional way to determine the pressure in a clamp cell during an experiment is to measure the unit-cell parameters of a material with a well characterized equation of state, such as NaCl, MgO, or Pb, mixed with the sample~\cite{Decker,Klotz_book}.
This method has several disadvantages:
i) The calibrant reduces much needed sample volume; ii) might absorb neutrons; and iii) it can react with the sample.
Also, such diffraction measurements are difficult with a neutron spectrometer, which may have insufficient $Q$-resolution, and require therefore an additional time consuming study using a neutron diffractometer.

The ruby fluorescence method~\cite{Forman284} is widely used for \emph{in-situ} pressure determination in a DAC, where optical access to the sample is available.
For the measurements, a small amount of ruby chip is added to the sample together with the pressure medium.
The pressure shift of a ruby $R_1$ laser-excited line is well calibrated at room temperature~\cite{Mao}.
The major challenge in ruby fluorescence measurements is that the temperature coefficient $d\lambda / dT$ is large, that is, the wavelength of the excitation depends not only on the pressure but also on the temperature~\cite{Vos,Syassen,Goncharov}.
Besides, the $R_1$ ruby line belongs to the $R_1 \& R_2$ doublet ($\sim$6942 and
6928 \AA, respectively) and the excited peaks are broadened under non-hydrostatic stress and change of temperature.
This leads to overlapping of the $R_1 \& R_2$  peaks, significantly reducing the accuracy of the pressure measurement.

Samarium-doped strontium tetraborate SrB$_4$O$_7$:Sm$^{2+}$ is an alternative material that largely avoids the limitations of ruby, owing to the very small temperature dependence of the wavelength of the $^7D_0 - ^5F_0$ excited line (0-0 hereafter), $d\lambda / dT \sim -0.001$~{\AA}/K comparing to $d\lambda / dT \sim 0.068$~{\AA}/K for the ruby excitation~\cite{Datchi,Lacam,Barnett}.
The 0-0 Sm$^{2+}$ excitation is a single line which is well isolated from the other fluorescence peaks~\cite{Lacam}, that makes it more suitable for accurate measurements.
The excitation is also little sensitive to non-hydrostatic stress~\cite{Datchi}.

The primary goal of our work was to build a reliable clamp pressure cell with an optical access to the sample space for accurate ($\Delta P < 0.1$~GPa) \emph{in-situ} pressure monitoring during inelastic neutron scattering (INS) experiments using the suite of the time-of-flight instruments on the Spallation Neutron Source (SNS) at Oak Ridge National Laboratory (ORNL)~\cite{Stone}, namely the cold neutron chopper spectrometer CNCS~\cite{CNCS1,CNCS2}, the hybrid spectrometer HYSPEC~\cite{HYSPEC}, the fine resolution Fermi chopper spectrometer SEQUOIA~\cite{SEQUOIA}, the Vibrational Spectrometer VISION~\cite{Seeger}, and the wide angular-range chopper spectrometer ARCS~\cite{ARCS}.
Our pressure cells can also be used for diffraction measurements at low temperature, but with limitation due to strong elastic background scattering.
To increase the reliability and accuracy of the pressure determination, we use two calibrants, ruby and Sr-borate, simultaneously.
The measuring system allows us to monitor pressure in the clamp cell during the loading process as well as during low temperature neutron scattering experiments.
We also present here a new calibration curve of the pressure-induced shift of 0-0 line of SrB$_4$O$_7$:Sm$^{2+}$ for moderate pressures $p \leqslant 2.0$~GPa.

\section{Experiment}

\begin{figure}[tbh!]
\centering
\includegraphics[width=1.0\columnwidth]{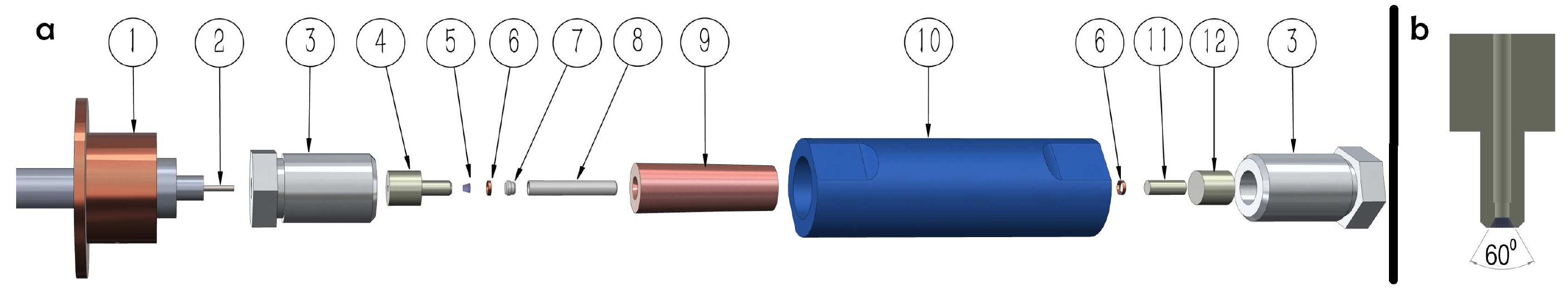}
  \caption{a) Schematic view and full assembly of the clamp cell used in the current study. (1) Cryostat sample stick; (2) Optical fiber; (3) Locking nut (Al-alloy); (4,11) Pistons (WC). The left piston has a hole for an optical fiber and (5) a diamond anvil; (6) Anti-extrusion rings (Cu); (7,8) Sample container (Teflon); (9) Inner sleeve (CuBe or NiCrAl alloy); (10) Main body (Al-alloy); (12) Support spacers (WC). b) Enlarged view of the piston (4).}
\label{pc}
\end{figure}

A schematic view of the clamp cell, that has been designed with the particular needs for neutron scattering experiments, is shown in Fig.~\ref{pc}(a).
The outer body, made from Al-alloy (7075 T651), is small enough (diameter 32~mm) to fit a $^3$He insert, that would allow to cool the sample down to $\sim$0.3~K.
The inner sleeve is made of either copper beryllium alloy~\cite{Klotz_book} or of Ni-Cr-Al alloy~\cite{Uwatoko,Fujiwara2007}, allowing pressures up to 1.6~GPa and 2.0~GPa, respectively.
The inner sleeve is slightly conical and pushed into the outer body to produce a radial support.

The piston (Fig.~\ref{pc}(b)) has a hole (1.8 ~mm diameter) in order to provide optical access to the sample space.
The piston is made of Bohler S390PM which is a micro-grain tool steel hardened to Rockwell Rc-67.
It is produced through powder metallurgy techniques and machined to tight tolerances.
The diamond seat is machined at a 60-degree included angle and honed to precisely fit the diamond.
The hole through the piston is sized to position the fiber optic probe for good optical access with the sample.
Installation of the diamond is accomplished by squarely seating the diamond and applying a small amount of optically clear Stycast epoxy around the top edge of the diamond where it contacts the piston.

The sample is placed in a teflon tube of $\sim{4.5}$~mm diameter and $\sim{15}$~mm length, allowing about 250~mm$^3$ of sample volume.
During the measurements the top and bottom edges of the pressure cell are covered with cadmium to reduce the background from residual scattering.
As a pressure transmitting medium we use either Fluorinert FC-770 or deuterated methanol.
Both fluids ensure quasi-hydrostatic conditions in a pressure range of our interest~\cite{Piermarini1973,Klotz2009} and we did not find any difference between them.
Samples to be measured can be either oriented single crystals or  poly-crystalline material.
All materials used in the construction of the pressure cell are nonmagnetic.
Both pressure cells (with either Cu-Be or Ni-Cr-Al sleeve) were successfully tested in a cryomagnet in fields up to 8~T and temperatures down to 1.7~K.

The ruby powder is commercially available.
The Sr-borate was prepared by reacting strontium carbonate with metaboric acid at high temperature.
Metaboric acid was prepared by heating boric acid in air for 24 hours at a temperature of $130-150^{\circ}$C,
3B(OH)$_3 \longrightarrow$~(BOH)$_3$O$_3 + $3H$_2$O.
In a typical synthesis 5 grams of boric acid were placed in a tall cylindrical alumina crucible heated in air in a convection oven.
The metaboric acid thus prepared was mixed with strontium carbonate and a small amount of samarium oxide corresponding to the desired level of doping, 3SrCO$_3 + 4$(BOH)$_3$O$_3 \longrightarrow$~3SrB$_4$O$_7 + $6H$_2$O$ + $3CO$_2$.
In a typical synthesis, 4.86~g of SrCO$_3$ (20~mmol; MW~147.63) were mixed with 3.51~g of (BOH)$_3$O$_3$ (27~mmol; MW~131.43).
After mixing and grinding the powders together, the samarium dopant was added and grinding was continued.
The resulting mixture was pressed into several pellets (1~cm diameter die; 6 tons).
The pellets were placed in a cylindrical alumina crucible (3~cm diameter $\times$ 5~cm height), and heated in a programmable muffle furnace.
The temperature was ramped from room temperature to $800^{\circ}$C over 60 minutes.
The furnace temperature was then held at $800^{\circ}$C for 6 hours.
The furnace temperature was then further raised to $850^{\circ}$C over a period of 30 minutes and held at $850^{\circ}$C for 12 hours.
The furnace temperature was then finally increased again to $880^{\circ}$C over 30 minutes and held at that temperature for another 6 hours, at which point the sample was cooled to room temperature over a period of 3 hours (about $5^{\circ}$C/min).
The solid mass recovered in the crucible was ground up and x-ray diffraction showed the expected strontium tetraborate structure.
An additional annealing step above $900^{\circ}$C (but below the $994^{\circ}$C melting point) after grinding improved the  crystallinity slightly.

The excitation light was provided by a 5320~{\AA} line of a LRS-0532 Diode-Pumped Solid-State (DPSS) Laser from Laserglow~\cite{Laser} with a maximum power of 200~mW.
For detection we use a HR4000 spectrometer from Ocean Optics~\cite{Ocean} calibrated for the wavelength $6800 - 7170$~{\AA} with Toshiba TCD1304AP linear CCD array (3648 pixels).
The optical fiber line is incorporated into the standard sample stick that allows us to monitor pressure in the cryostat, the cryomagnet or the closed cycle refrigerator (CCR).

We used a DAC for the calibration of the pressure-induced shift of the 0-0 line of SrB$_4$O$_7$:Sm$^{2+}$.
The DAC allowed us to cover a pressure range exceeding the clamp cell limit in order to ensure the accuracy of the obtained calibration curve.

\section{Results and Discussion}

\subsection{Empty cell background}

\begin{figure}[t!]
\centering
\includegraphics[width=0.95\columnwidth]{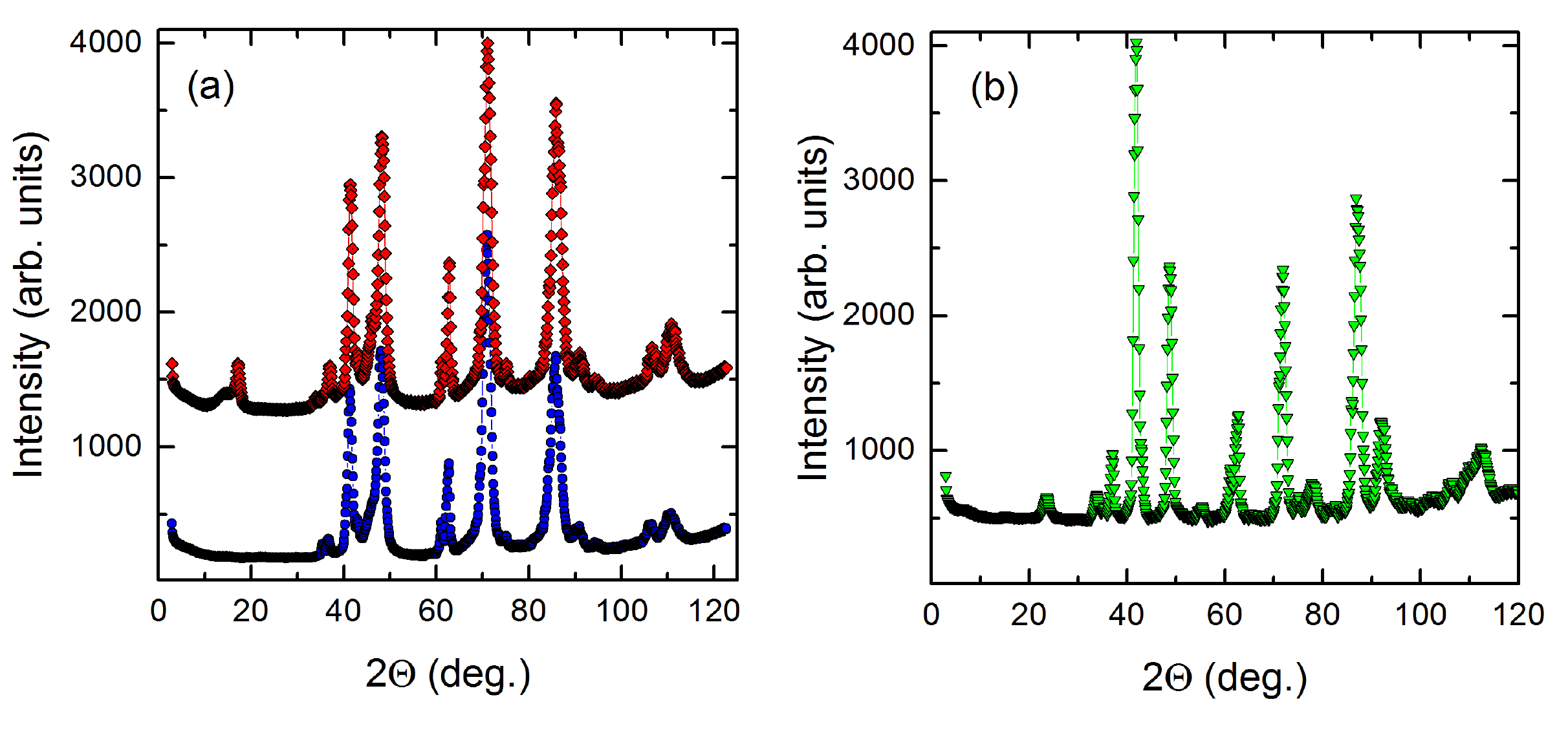}
  \caption{The room temperature neutron diffraction patterns of a) the empty cell with Cu-Be sleeve (the blue bottom pattern) and the empty cell loaded with a Teflon container and fluorinert (the red upper pattern); b) The empty cell with Ni-Cr-Al sleeve. In (a) the diffraction patterns are shifted for clarity. Note the low angle scattering, $2\Theta\sim{15}$ degrees, from the  pressure transmitting medium.}
\label{diffr}
\end{figure}

The materials used to construct the pressure cell absorb and scatter neutrons (elastically and inelastically; coherently and incoherently), reducing the incident neutron flux on the sample and contributing to the scattered signal.
This beam attenuation and background scattering increases counting time, reduces signal to noise ratio and produces overlapping peaks.
In order to estimate the background and scattering profile of the sample environment, we performed both neutron diffraction and INS measurements of the empty pressure cells.
Neutron diffraction test measurements were done using the wide-angle neutron diffractometer WAND$^2$~\cite{Wand} at the High Flux Isotope Reactor (HFIR) reactor at ORNL.
An incident neutron beam with a wavelength of 1.4827~{\AA} (37.2~meV) was selected with a Ge (113) monochromator.
Fig.~\ref{diffr} shows the room temperature neutron diffraction patterns of the empty pressure cell with (a) Cu-Be and (b) Ni-Cr-Al inner sleeve.
Strong Bragg peaks at scattering angles $2\Theta > 40$ degrees make diffraction measurements and the Rietveld refinement challenging.
However, the low-angle background 2$\Theta < 40$ degrees is almost free of elastic reflections, except scattering from the pressure transmitting medium.
Therefore, the pressure cell can be used for structural studies of magnetic materials provided the empty cell is measured under identical experimental conditions and the scattering signal subtracted.
We are planning to build a clamp cell specifically for  neutron diffraction, made of titanium-zirconium alloy with null scattering composition~\cite{Klotz_book}, which will not produce Bragg reflections in a neutron beam.

A study of neutron beam attenuation by the pressure cells and inelastic scattering background measurements were performed at  CNCS.
The aluminum alloys of the pressure body have high neutron transparency due to the small absorption and relatively small incoherent and coherent cross sections of Al ($\sigma_{\mathrm{tot}}=1.503$~barn, $\sigma_{\mathrm{abs}}=0.231$~barn at $E_i=25.5$~meV)~\cite{Sears}.
It is Ni ($\sigma_{\mathrm{tot}}=18.5$~barn, $\sigma_{\mathrm{abs}}=4.49$~barn) and Cu ($\sigma_{\mathrm{tot}}=8.03$~barn, $\sigma_{\mathrm{abs}}=3.78$~barn) that gives the major contribution to the beam attenuation.
To determine the neutron transmission of the pressure cell we measured the Bragg reflections of a powder sample of the yttrium iron garnet Y$_3$Fe$_5$O$_{12}$ (YIG) outside and in the pressure cell at room temperature.
\begin{figure}[tbh!]
\centering
\includegraphics[width=0.4\columnwidth]{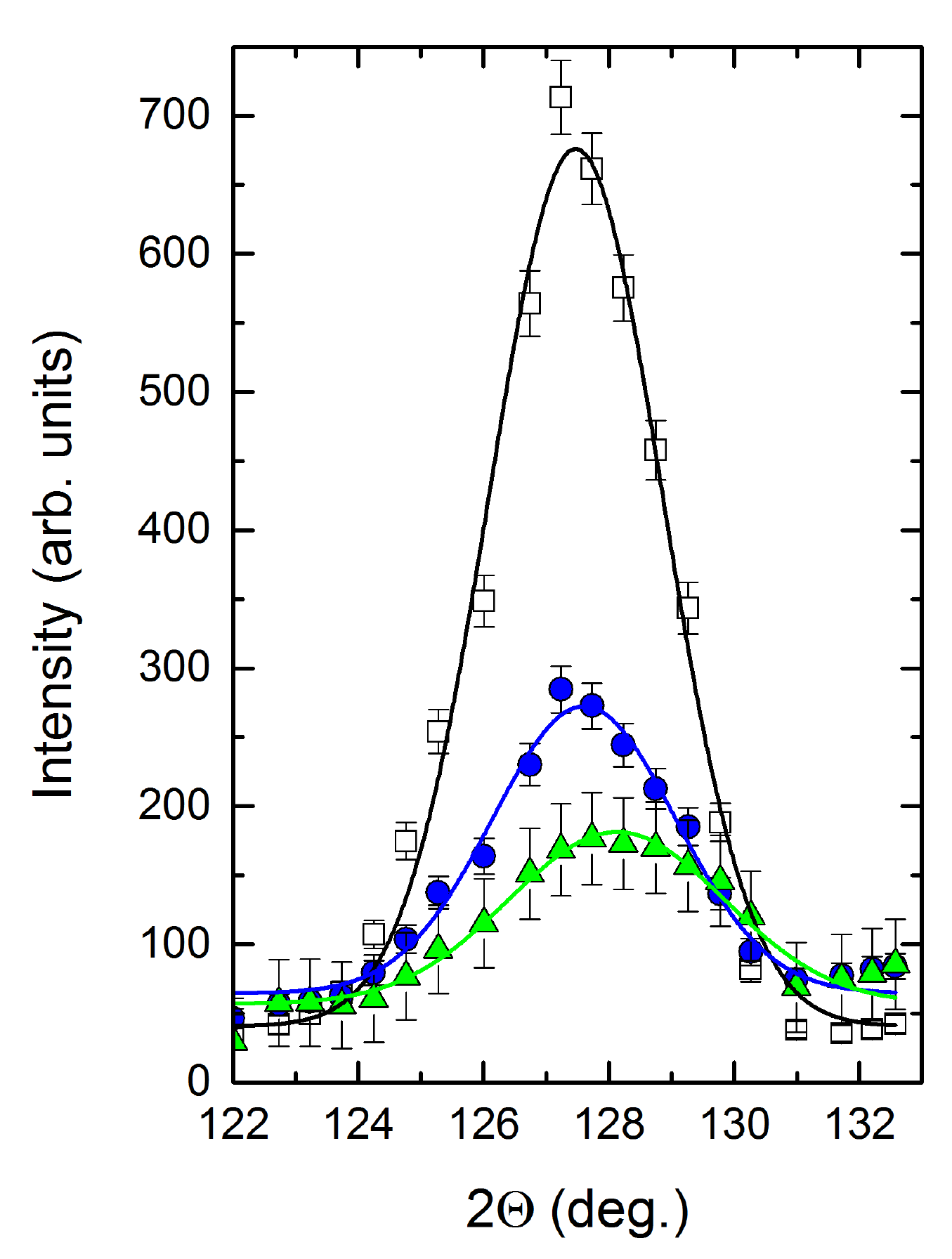}
  \caption{Integrated intensities of the (420) Bragg reflection of the YIG polycrystalline sample inside and outside of the pressure cell, that characterized the neutron transmission.  The sample was measured without the cell (open squares) and inside the cell with Cu-Be (circles) and Ni-Cr-Al sleeve (triangles) at the $E_i=3.3$~meV ($\lambda=4.96$~{\AA}). Solid lines are the Gaussian fit of the data. The pressure cell background was subtracted.}
\label{diffr_peak}
\end{figure}
Figure~\ref{diffr_peak} compares the integrated intensities of the (420) Bragg peak of the YIG sample, measured without and in the cells with  Ni-Cr-Al and  Cu-Be sleeves.
At the incident energy $E_i=3.3$~meV ($\lambda=4.96$~{\AA}), the ratio of the integrated intensity for the Cu-Be (Ni-Cr-Al) cell turns out to be $I_{\textrm{Cu-Be}}$/$I_{\textrm{no~cell}} = 0.35$ ($I_{\textrm{Ni-Cr-Al}}$/$I_{\textrm{no~cell}} = 0.24$), indicating a reasonable neutron transmission for both pressure cells.

\begin{figure}[tbh!]
\centering
\includegraphics[width=0.95\columnwidth]{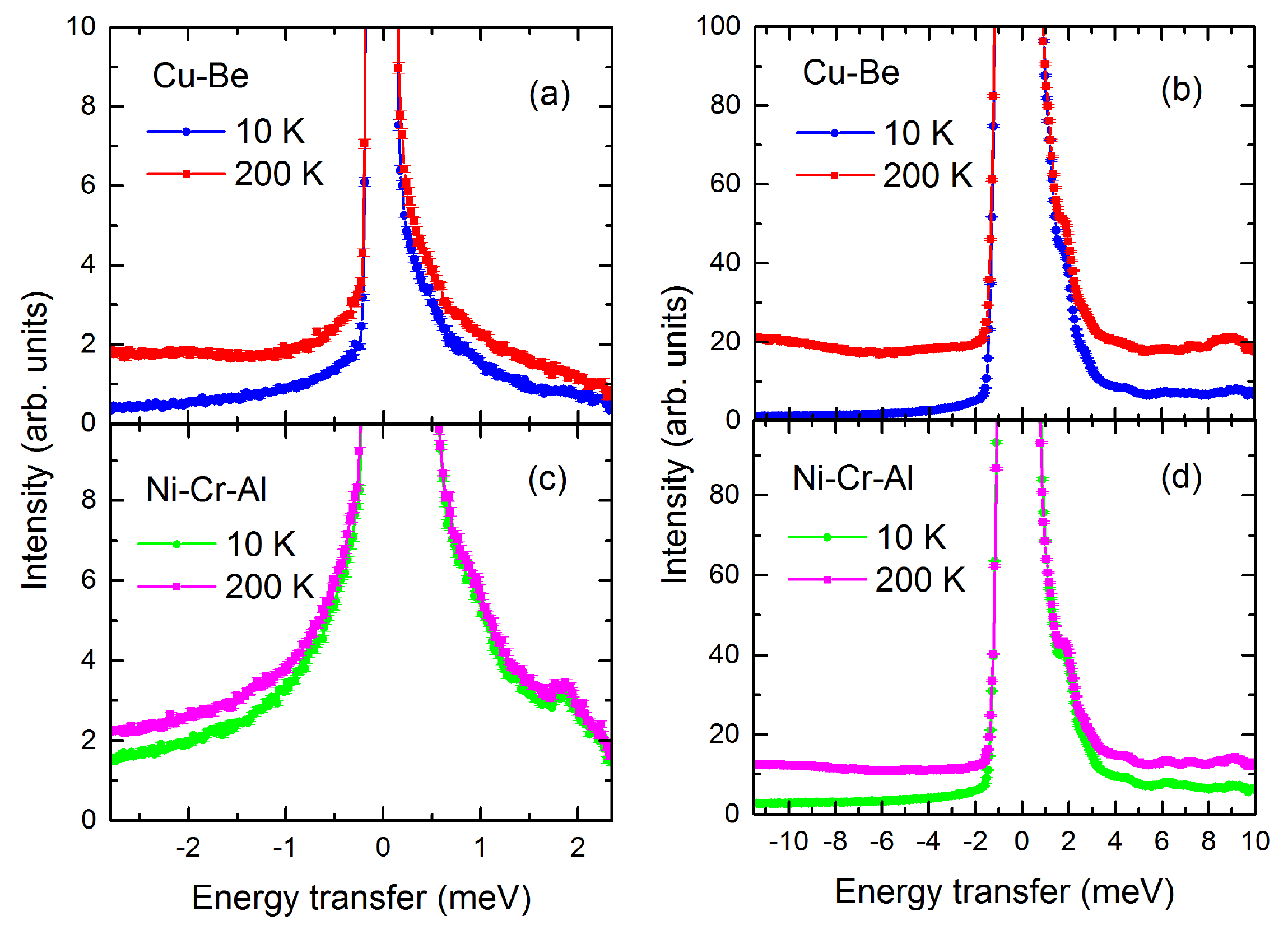}
  \caption{The INS spectra measured from the empty pressure cells with Cu-Be sleeve (a,b) and Ni-Cr-Al sleeve (c,d) at temperatures $T=10$ and 200~K. The spectra obtained with the incident energy $E_i=3.3$~meV (a,c) and $E_i=12.0$~meV (b,d) by integration in a low- $1.0<Q<1.5$~\AA$^{-1}$ and high-$Q$ $3.0<Q<3.5$~\AA$^{-1}$ range, respectively.}
\label{ins}
\end{figure}

Since experiments with magnetic materials often require cryogenic temperatures, we carried out  background measurements at two temperatures $T=10$ and 200~K.
Figure~\ref{ins} summarizes the INS background of the two empty cells with Cu-Be and Ni-Cr-Al sleeves for two incident neutron energies $E_i=3.3$~meV ($\lambda=4.96$~\AA) and $E_i=12.0$~meV ($\lambda=2.61$~\AA) often used in magnon and phonon measurements.
We conclude that the inelastic background is rather smooth and strongly dependent on the temperature.
The clear asymmetry of the elastic line is instrumental and also due to multiple scattering.

\subsection{Pressure calibration at room temperature}

\begin{figure}[tbh!]
\centering
\includegraphics[width=0.7\columnwidth]{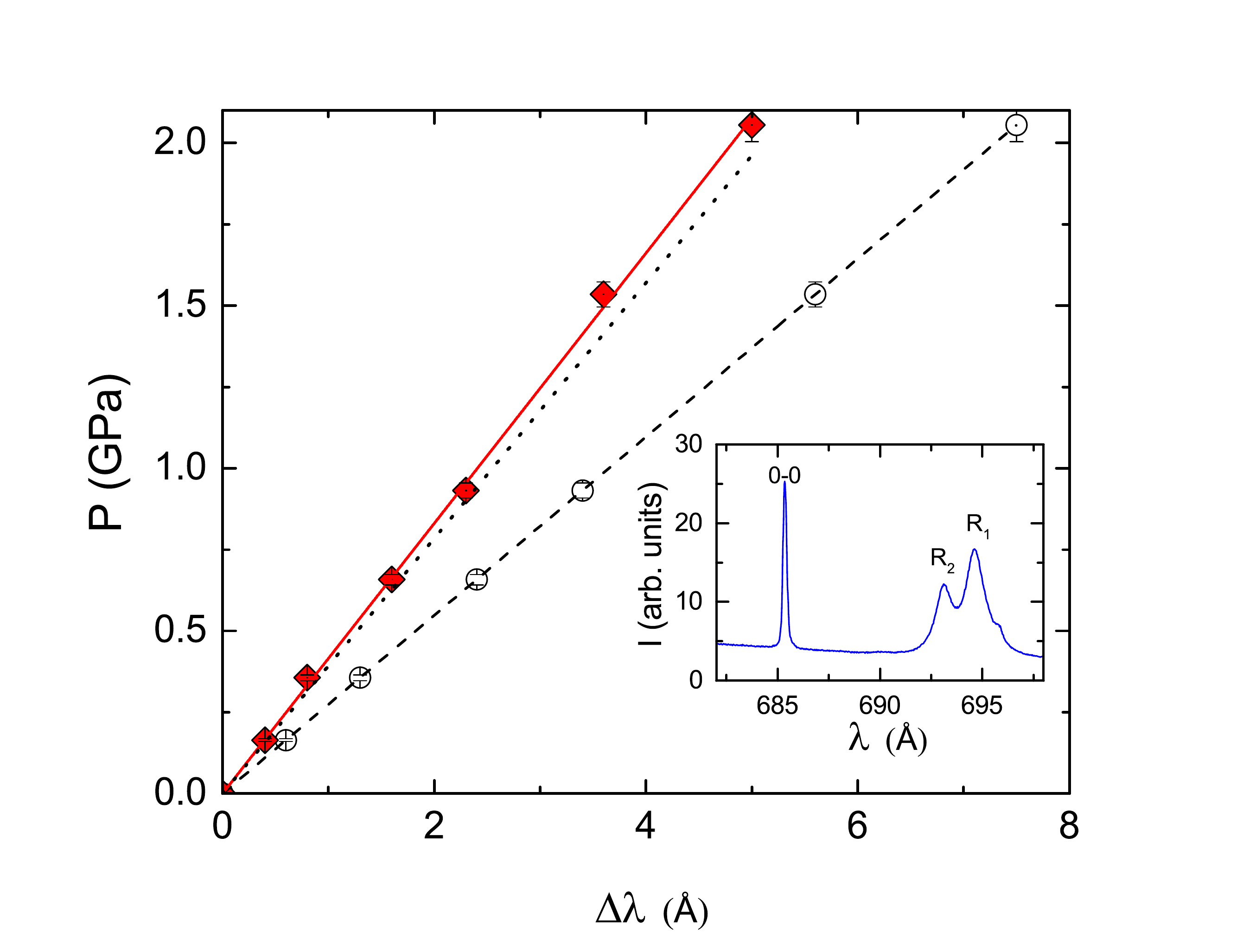}
  \caption{Calibration of the Sr-borate 0-0 line wavelength shift with pressure at room temperature. The circles and squares are the experimental data for the $R_1$ and the 0-0 excitations, respectively. The solid red line is a linear fit of the 0-0 line shift $P=\Delta \lambda / x$ with $x=2.41$. The dashed line is a linear equation for the $R_1$ ruby excitation, $\Delta \lambda_{R_1} /dP =3.65$~\AA/GPa~\cite{Datchi,Leger}. The dotted line is the linear extrapolation from high pressure data obtained by Lacam {\it{et al.}}~\cite{Lacam}, $\Delta \lambda_{R_1} /dP =2.55$~\AA/GPa. The inset shows a typical luminescence spectrum of 0-0 Sm$^{2+}$ and $R_1\&R_2$ ruby excitations.
}
\label{calibration}
\end{figure}

The pressure shift of the ruby excitation line $\Delta \lambda_{R_2}$ at room temperature is well calibrated in a wide pressure range~\cite{Piermarini,Mao} by the following equation:
\begin{equation}
P = 248.4\left[\left(\frac{\lambda_{R_1}(P)}{\lambda_{R_1}(0)}\right)^{7.665} - 1\right],
\label{ruby}
\end{equation}
where $\lambda_{R_2}(0) = 6942.2$~{\AA} is a wavelength at ambient pressure, $P$ is expressed in GPa and $\lambda$ in {\AA}.
Note, that the nonlinearity of the variation of $\lambda_{R_1}$ as a function of pressure is negligible for $P<2.0$~GPa.
We also assume that the slope of the temperature dependence of the ruby R$_1$ line is pressure independent in our pressure range~\cite{Goncharov}.
Therefore, for all practical use we can adopt the following linear equation, $\Delta \lambda_{R_1} /dP =3.65 \pm 0.05$~{\AA}/GPa~\cite{Datchi,Leger}, which is undistinguished from Equation~\ref{ruby} (shown as a dashed line in Figure~\ref{calibration}).

The calibration of the samarium-doped strontium tetraborate fluorescence was conducted against the ruby fluorescence in a DAC.
Ruby chips and Sr-borate powder were placed on the top anvil of a Boehler plate DAC~\cite{Boehler2006} (650~$\mu$m diameter culets, 301 stainless steel gasket with chamber diameter 300~$\mu$m and height 75~$\mu$m).
The cell was then filled with deuterated glycerin as hydrostatic pressure transmitting medium~\cite{Klotz2012}.
The pressure was gradually increased using the plate DAC gears.
The fluorescence was measured using the SNAP Raman stand equipped with a 532~nm laser and a 1800~mm$^{-1}$ grating.
Spectra were typically acquired within 0.5~s.
A spot in the cell was identified that allowed for good quality detection of the ruby and Sr-borate excitations simultaneously (see inset in Figure~\ref{calibration}).
The same spot was used for all measurements.
Ruby as well as Sr-borate shifts are given relative to a measurement acquired prior to compression.
We found that for  pressures $P<2.0$~GPa the wavelength pressure shift of the $0-0$ line is well fitted by a $\Delta \lambda_{0-0} /dP =2.41(1)$~{\AA}/GPa linear relation, see Figure~\ref{calibration}.
Note, that the linear coefficient we determined in this work is slightly different from the one obtained by Lacam {\it{et al.}}~\cite{Lacam} and Datchi {\it{et al.}}~\cite{Datchi2007}.

\subsection{Temperature effect on the pressure}

\begin{figure}[tbh!]
\centering
\includegraphics[width=1.0\columnwidth]{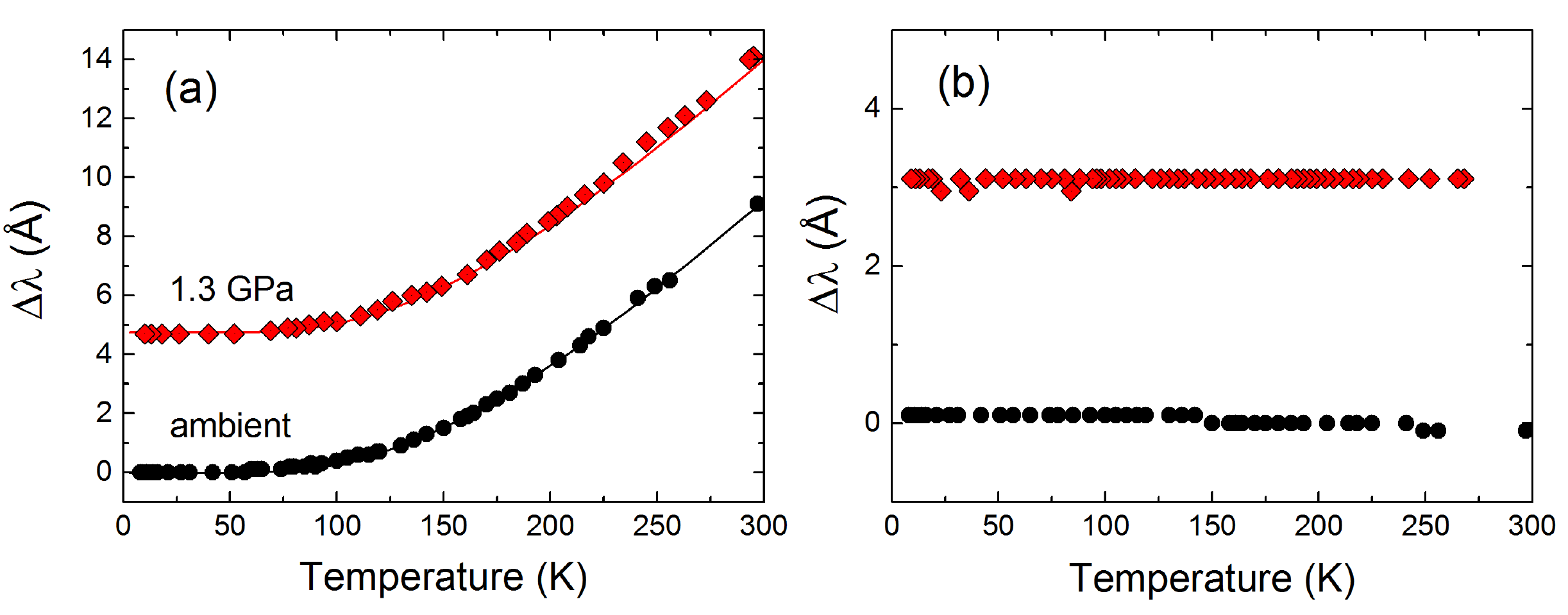}
  \caption{A shift of the R$_1$ (a) and 0-0 (b) emission lines as a function of temperature at ambient pressure (black circles) and $P=1.3$~GPa (red squares). Solid lines are a fit of the experimental data to Equation~\ref{tdep}, see text.}
\label{Tempdep}
\end{figure}

The temperature effect on the R-line emission of ruby has been known for decades~\cite{McCumber,Vos}.
A detailed analysis of the physical properties of ruby and of the origin of the R-line shift with temperature and pressure has been performed by K.~Syassen~\cite{Syassen}.
We used the pressure cell with CuBe sleeve for the temperature dependent measurements.
For the pressure application the sample space of the cell was filled with the pressure medium and a mixture of ruby chips and SrB$_4$O$_7$:Sm$^{2+}$ powder.
Pressure was applied by a hydraulic press and monitored during the load.
The deviation of the real pressure compared to the nominal pressure is about 20-30\%.
For the temperature measurements the pressure cell was mounted on the cold plate of a standard Close Cycle Refrigerator.
Fig.~\ref{Tempdep} shows the temperature effect on both, $R_1$ and 0-0 lines, at ambient pressure as well as at $P=1.3$~GPa.
With decreasing temperature, the $R_1$-line shifts to lower wavelength while the 0-0 line does not shift.
Our data for the $R_1$-line agrees well with an earlier published model~\cite{Syassen}, where the temperature dependence of the shift is fitted by an analytical expression
\begin{equation}
\nu(T) = \nu_0 - \frac{\alpha_{\nu}}{exp{(\Theta/T)}-1},
\label{tdep}
\end{equation}
with parameters $\nu_0=14423.4$~cm$^{-1}$ (that correspond to $\lambda_0=6933.2$~{\AA}), $\alpha_{\nu}=76.6$~cm$^{-1}$ and $\Theta=482$~K.
It follows from Fig.~\ref{Tempdep}(b) that the pressure drop at $P=1.3$~GPa is negligible between ambient temperature and 10~K.

\section{Conclusions}

To conclude, we fabricated and tested nonmagnetic clamp pressure cells with Cu-Be ($P\leqslant{1.6}$~GPa) and Ni-Cr-Al ($P\leqslant{2.0}$~GPa) inner sleeves, which are suitable for low temperature neutron diffraction and inelastic neutron scattering experiments including measurements in external magnetic fields.

Summarizing our neutron scattering measurements of the pressure cell background, we observe significant coherent elastic scattering, as expected.
This limits an applicability of the cells for diffraction measurements.
However, the refinement of the magnetic structure transformation, based mainly on the low-angle magnetic Bragg peaks, is often possible  when the empty cell is measured under identical conditions and its signal is subtracted.
The neutron transmission is reasonably good for a large energy interval down to $E_{i}\sim{3}$~meV.
The inelastic background scattering, although high, is smooth at low energy transfer up to $\hbar\omega<10$~meV.
There are a large number of scientific subjects, which fit well to these parameters, for example, low-energy phonons, magnons and crystal-field excitations, as well as quantum phenomena such as quantum tunneling and quantum spin fluctuations.
In general, measurements of the empty cell and a subtraction from the experimental data is highly recommended for both types, diffraction and INS experiments.

The pressure cells have an optical access to the sample space for fluorescence measurements.
The temperature independent linear pressure shift $\Delta \lambda_{0-0} /dP =2.41(1)$~\AA/GPa of the 0-0 fluorescent line of SrB$_4$O$_7$:Sm$^{2+}$ allows an accurate \emph{in-situ} pressure determination in the entire temperature range from room temperature down to 1.5~K.
Cross-checking the 0-0 excitation versus the ruby line increases the reliability of the measurements.
The pressure applied to the sample can be determined with an accuracy better than $\pm{0.05}$~GPa.

Currently this equipment is accessible at ORNL's neutron scattering research facilities - the Spallation Neutron Source and the High Flux Isotope Reactor, through ORNL's general user program~\cite{ORNL}.

\section*{Acknowledgement}

This research used resources at the Spallation Neutron Source and the High Flux Isotope Reactor, a DOE Office of Science User Facility operated by the Oak Ridge National Laboratory.

\section*{Disclosure statement}

No potential conflict of interest was reported by the authors.



\begin{thebibliography}{10}
\providecommand{\url}[1]{\normalfont{#1}}
\providecommand{\urlprefix}{Available from: }

\bibitem{Boehler}
Boehler~R, Guthrie~M, Molaison~J, et~al. {Large-volume diamond cells for
  neutron diffraction above 90~GPa}. High Press Res.
  2013;\hspace{0pt}33(3):546--554.

\bibitem{Guthrie10552}
Guthrie~M, Boehler~R, Tulk~CA, et~al. Neutron diffraction observations of
  interstitial protons in dense ice. Proc Natl Acad Sci U S A.
  2013;\hspace{0pt}110(26):10552--10556.

\bibitem{Boehler2017}
Boehler~R, Molaison~JJ, Haberl~B. {Novel diamond cells for neutron diffraction
  using multi-carat CVD anvils}. Rev Sci Instrum.
  2017;\hspace{0pt}88(8):083905.

\bibitem{Tafti}
Tafti~FF, Juneau-Fecteau~A, Delage~ME, et~al. {Sudden reversal in the pressure
  dependence of $T_{\mathrm{c}}$ in the iron-based superconductor
  KFe$_2$As$_2$}. Nature Physics. 2013;\hspace{0pt}9:349--352.

\bibitem{Chen}
Chen~Y, Jiang~WB, Guo~CY, et~al. {Reemergent Superconductivity and Avoided
  Quantum Criticality in Cd-Doped ${\mathrm{CeIrIn}}_{5}$ under Pressure}. Phys
  Rev Lett. 2015;\hspace{0pt}114:146403.

\bibitem{Luo13520}
Luo~Y, Ronning~F, Wakeham~N, et~al. {Pressure-tuned quantum criticality in the
  antiferromagnetic Kondo semimetal CeNi$_{2-\delta}$As$_2$}. Proc Natl Acad
  Sci U S A. 2015;\hspace{0pt}112(44):13520--13524.

\bibitem{Fujiwara}
Fujiwara~N, Kawaguchi~N, IImura~S, et~al. {Quantum phase transition under
  pressure in the heavily hydrogen-doped iron-based superconductor LaFeAsO}.
  Phys Rev B. 2017;\hspace{0pt}96:140507.

\bibitem{Li}
Li~W, Wei~XY, Zhu~JX, et~al. {Pressure-induced topological quantum phase
  transition in Sb${}_{2}$Se${}_{3}$}. Phys Rev B. 2014;\hspace{0pt}89:035101.

\bibitem{Cai}
Cai~PL, Hu~J, He~LP, et~al. {Drastic Pressure Effect on the Extremely Large
  Magnetoresistance in ${\mathrm{WTe}}_{2}$: Quantum Oscillation Study}. Phys
  Rev Lett. 2015;\hspace{0pt}115:057202.

\bibitem{Cheng1670}
Cheng~J, Kweon~KE, Larregola~SA, et~al. {Charge disproportionation and the
  pressure-induced insulator{\textendash}metal transition in cubic perovskite
  PbCrO$_3$}. Proc Natl Acad Sci U S A. 2015;\hspace{0pt}112(6):1670--1674.

\bibitem{Kengo}
Oka~K, Azuma~M, Chen~W, et~al. {Pressure-Induced Spin-State Transition in
  BiCoO$_3$}. J Am Chem Soc. 2010;\hspace{0pt}132(27):9438--9443.

\bibitem{Klotz_book}
Klotz~S. {Techniques in High Pressure Neutron Scattering}. CRC Press; 2013.

\bibitem{Decker}
Decker~DL. {High-Pressure Equation of State for NaCl, KCl, and CsCl}. J Appl
  Phys. 1971;\hspace{0pt}42(8):3239--3244.

\bibitem{Forman284}
Forman~RA, Piermarini~GJ, Barnett~JD, et~al. {Pressure Measurement Made by the
  Utilization of Ruby Sharp-Line Luminescence}. Science.
  1972;\hspace{0pt}176(4032):284--285.

\bibitem{Mao}
Mao~HK, Xu~J, Bell~PM. {Calibration of the ruby pressure gauge to 800 kbar
  under quasi-hydrostatic conditions}. J Geophys Res Solid Earth.
  1986;\hspace{0pt}91(B5):4673--4676.

\bibitem{Vos}
Vos~WL, Schouten~JA. On the temperature correction to the ruby pressure scale.
  J Appl Phys. 1991;\hspace{0pt}69(9):6744--6746.

\bibitem{Syassen}
Syassen~K. {Ruby under pressure}. High Press Res.
  2008;\hspace{0pt}28(2):75--126.

\bibitem{Goncharov}
Goncharov~AF, Zaug~JM, Crowhurst~JC, et~al. {Optical calibration of pressure
  sensors for high pressures and temperatures}. J Appl Phys.
  2005;\hspace{0pt}97(9):094917.

\bibitem{Datchi}
Datchi~F, LeToullec~R, Loubeyre~P. {Improved calibration of the
  SrB$_4$O$_7$:Sm$^{2+}$ optical pressure gauge: Advantages at very high
  pressures and high temperatures}. J Appl Phys.
  1997;\hspace{0pt}81(8):3333--3339.

\bibitem{Lacam}
Lacam~A, Chateau~C. {High-pressure measurements at moderate temperatures in a
  diamond anvil cell with a new optical sensor: SrB$_4$O$_7$:Sm$^{2+}$}. J Appl
  Phys. 1989;\hspace{0pt}66(1):366--372.

\bibitem{Barnett}
Barnett~JD, Block~S, Piermarini~GJ. {An Optical Fluorescence System for
  Quantitative Pressure Measurement in the Diamond-Anvil Cell}. Rev Sci
  Instrum. 1973;\hspace{0pt}44(1):1--9.

\bibitem{Stone}
Stone~MB, Niedziela~JL, Abernathy~DL, et~al. {A comparison of four direct
  geometry time-of-flight spectrometers at the Spallation Neutron Source}. Rev
  Sci Instr. 2014;\hspace{0pt}85(4):045113.

\bibitem{CNCS1}
Ehlers~G, Podlesnyak~A, Niedziela~JL, et~al. {The new cold neutron chopper
  spectrometer at the Spallation Neutron Source: design and performance}. Rev
  Sci Instrum. 2011;\hspace{0pt}82:085108.

\bibitem{CNCS2}
Ehlers~G, Podlesnyak~A, Kolesnikov~AI. {The cold neutron chopper spectrometer
  at the Spallation Neutron Source - A review of the first 8 years of
  operation}. Rev Sci Instrum. 2016;\hspace{0pt}87:093902.

\bibitem{HYSPEC}
{Winn, B}, {Filges, U}, {Garlea, V O}, et~al. {Recent progress on HYSPEC, and
  its polarization analysis capabilities}. EPJ Web Conf.
  2015;\hspace{0pt}83:03017.

\bibitem{SEQUOIA}
Granroth~G, Kolesnikov~A, Sherline~T, et~al. {SEQUOIA: A newly operating
  chopper spectrometer at the SNS}. J Phys: Conf Ser.
  2010;\hspace{0pt}251(1):012058.

\bibitem{Seeger}
Seeger~PA, Daemen~LL, Larese~JZ. {Resolution of VISION, a Crystal-Analyzer
  Spectrometer}. Nucl Instr Meth A. 2009;\hspace{0pt}604:719--728.

\bibitem{ARCS}
Abernathy~DL, Stone~MB, Loguillo~MJ, et~al. {Design and operation of the wide
  angular-range chopper spectrometer ARCS at the Spallation Neutron Source}.
  Rev Sci Instrum. 2012;\hspace{0pt}83(1):015114.

\bibitem{Uwatoko}
Uwatoko~Y, Todo~S, Ueda~K, et~al. {Material properties of Ni–Cr–Al alloy
  and design of a 4~GPa class non-magnetic high-pressure cell}. J Phys: Condens
  Matter. 2002;\hspace{0pt}14(44):11291.

\bibitem{Fujiwara2007}
Fujiwara~N, Matsumoto~T, Nakazawab~K, et~al. Fabrication and efficiency
  evaluation of a hybrid nicral pressure cell up to 4~gpa. Rev Sci Instrum.
  2007;\hspace{0pt}78(7):073905.

\bibitem{Piermarini1973}
Piermarini~GJ, Block~S, Barnett~J. Hydrostatic limits in liquids and solids to
  100 kbar. J Appl Phys. 1973;\hspace{0pt}44(12):5377--5382.

\bibitem{Klotz2009}
Klotz~S, Chervin~JC, Munsch~P, et~al. {Hydrostatic limits of 11 pressure
  transmitting media}. J Appl Phys. 2009;\hspace{0pt}42(7):075413.

\bibitem{Laser}
 {LASERGLOW TECHNOLOGIES, 873 St. Clair Ave West, Toronto, ON, Canada},
  M6C1C4.

\bibitem{Ocean}
 {OCEAN OPTICS, 8060 Bryan Dairy Rd, Largo FL 33777}, USA.

\bibitem{Wand}
Frontzek~M, Andrews~K, Jones~A, et~al. {The Wide Angle Neutron Diffractometer
  squared (WAND$^2$) - Possibilities and Future}. Physica B: Condens Matter.
  2017;\hspace{0pt}in press.

\bibitem{Sears}
Sears~VF. Neutron scattering lengths and cross sections. Neutron News.
  1992;\hspace{0pt}3(3):26--37.

\bibitem{Piermarini}
Piermarini~GJ, Block~S, Barnett~JD, et~al. {Calibration of the pressure
  dependence of the R$_1$ ruby fluorescence line to 195 kbar}. J Geophys Res
  Solid Earth. 1975;\hspace{0pt}46(6):2774--2780.

\bibitem{Leger}
Leger~JM, Chateau~C, Lacam~A. {SrB$_4$O$_7$:Sm$^{2+}$ pressure optical sensor:
  Investigations in the megabar range}. J Appl Phys.
  1990;\hspace{0pt}68(5):2351--2354.

\bibitem{Boehler2006}
Boehler, Reinhard. New diamond cell for single-crystal x-ray diffraction. Rev
  Sci Instrum. 2006;\hspace{0pt}77(11):115103.

\bibitem{Klotz2012}
Klotz~S, Takemura~K, Str\"assle~T, et~al. {Freezing of glycerol–water
  mixtures under pressure}. J Phys: Condens Matter. 2012;\hspace{0pt}24:325103.

\bibitem{Datchi2007}
Datchi~F, Dewaele~A, Loubeyre~P, et~al. Optical pressure sensors for
  high-pressure–high-temperature studies in a diamond anvil cell. High Press
  Res. 2007;\hspace{0pt}27(4):447--463.

\bibitem{McCumber}
McCumber~DE, Sturge~MD. {Linewidth and Temperature Shift of the R Lines in
  Ruby}. J Appl Phys. 1963;\hspace{0pt}34(6):1682--1684.

\bibitem{ORNL}
 \url{https://neutrons.ornl.gov/users}.

\end{thebibliography}

\end{document}